\documentclass{emulateapj}
\usepackage{url,graphicx,amssymb}
\usepackage{subfigure,lscape}

\long\def\symbolfootnote[#1]#2{\begingroup%
\def\thefootnote{\fnsymbol{footnote}}\footnote[#1]{#2}\endgroup}

\newcommand\mq{1E 1740.7-2942}

\begin{document}

\shorttitle{Suzaku observations of \mq}
\shortauthors{Reynolds et al.}

\title
{Suzaku observations of the Galactic center microquasar \mq}
\author{Mark T. Reynolds\altaffilmark{1}, Jon M. Miller\altaffilmark{1}}  
\email{markrey@umich.edu}

\altaffiltext{1}{Department of Astronomy, University of Michigan, 500 Church
  Street, Ann Arbor, MI 48109} 

\begin{abstract}
We present two \textit{Suzaku} observations of the Galactic center
microquasar \mq~separated by approximately 700 days. The source was observed
on both occasions after a transition to the spectrally hard
state. Significant emission from \mq~ is detected out to an energy of 300
keV, with no spectral break or turnover evident in the data. We tentatively
measure a lower limit to the cut-off energy of $\sim$ 380 keV.  The spectra
are found to be consistent with a Comptonized corona on both occasions,
where the high energy emission is consistent with a hard power-law ($\rm
\Gamma \sim 1.8$) with a significant contribution from an accretion disc
with a temperature of $\sim$ 0.4 keV at soft X-ray energies.

The measured value for the inner radius of the accretion disc is found to be
inconsistent with the picture whereby the disc is truncated at large radii
in the low-hard state and instead favours a radius close to the ISCO ($\rm
R_{in} \sim 10 - 20~R_g$). 
\end{abstract}
 
\keywords{accretion, accretion disks --- black hole physics --- X-rays:
  binaries --- X-rays: individual (\mq)} 

\maketitle
\section{Introduction}
Since the discovery of apparently super-luminal jets from the X-ray binary
GRS 1915+105 \citep{b12}, the Galactic microquasars have assumed a position
of critical importance in our efforts to understand accretion physics and
relativistic jet production \citep{b40}.  \mq~ was discovered by the
\textit{Einstein} satellite \citep{b1}. Subsequent observations revealed
\mq~to be the dominant source of hard X-rays ($\rm > 20~keV$) in the
direction of the Galactic center \citep{b2,b3}, where the source is located
approximately 50 arcmin from Sgr A$\rm ^*$.  The microquasar nature of
\mq~was discovered upon the observation of a double sided radio jet
consistent with the X-ray position \citep{b4}.  Further VLA observations
showed this radio source to be highly variable \citep{b5}.

Since its discovery \mq~has been observed on numerous occasion at X-ray
wavelengths. The column density towards this source is high given the
proximity to the Galactic center and has been measured by \textit{Chandra}
to be $\rm \sim 1 \times 10^{23}~cm^{-2}$ \citep{b10}. Here, the spectrum
was found to be consistent with a power-law ($\rm \Gamma \sim 1.4$).
\textit{INTEGRAL} low-hard state observations have detected \mq~up to
energies of $\sim$ 600 keV, where the spectrum is found to be consistent
with a power-law ($\rm \Gamma \sim 1.6$) up to 200 keV, with an additional
component required at higher energies \citep{b8}, see also \citet{b9} for
earlier \textit{INTEGRAL/RXTE} observations. In the high-soft state,
the hard X-ray flux decreases significantly, dropping below the
\textit{INTEGRAL} detection limit at energies $\rm >$ 50 keV
\citep{b8}. \citet{b11} reported on 5 years of \textit{RXTE} monitoring,
where they discovered a modulation with a period 12.73$\rm \pm$0.05 days
which is attributed to the orbital period of the binary, in addition to a
possible super-orbital modulation with a period of $\sim$ 600 days.

As the extinction at optical wavelengths is prohibitive ($\rm A_v \sim 50$),
counterpart searches are required to take place in the infrared. While a
number of candidate counterparts have been identified \citep{b6,b7}, no
variability has been observed from these, rendering them unlikely to be the
actual counterpart. The current upper limit for the K$\rm_s$-band magnitude
of the counterpart is $\geq$ 19.9 at the 95\% confidence level. This is
equivalent to a secondary spectral type of O or B if on the main sequence or
K if a giant, at an assumed distance of 8.5 kpc.

In this paper, we describe observations undertaken with the \textit{Suzaku}
X-ray observatory, while \mq~was in the low-hard state. In \S2, we
describe the observations and extraction of source spectra. We proceed to
analyze the data in \S3. In \S4, these results are compared to observations
of other microquasars in the hard state, and finally our conclusions are
presented in \S5.

\section{Observations}
\begin{figure}
\begin{center}
\includegraphics[height=0.34\textheight,width=0.3\textwidth,angle=-90]{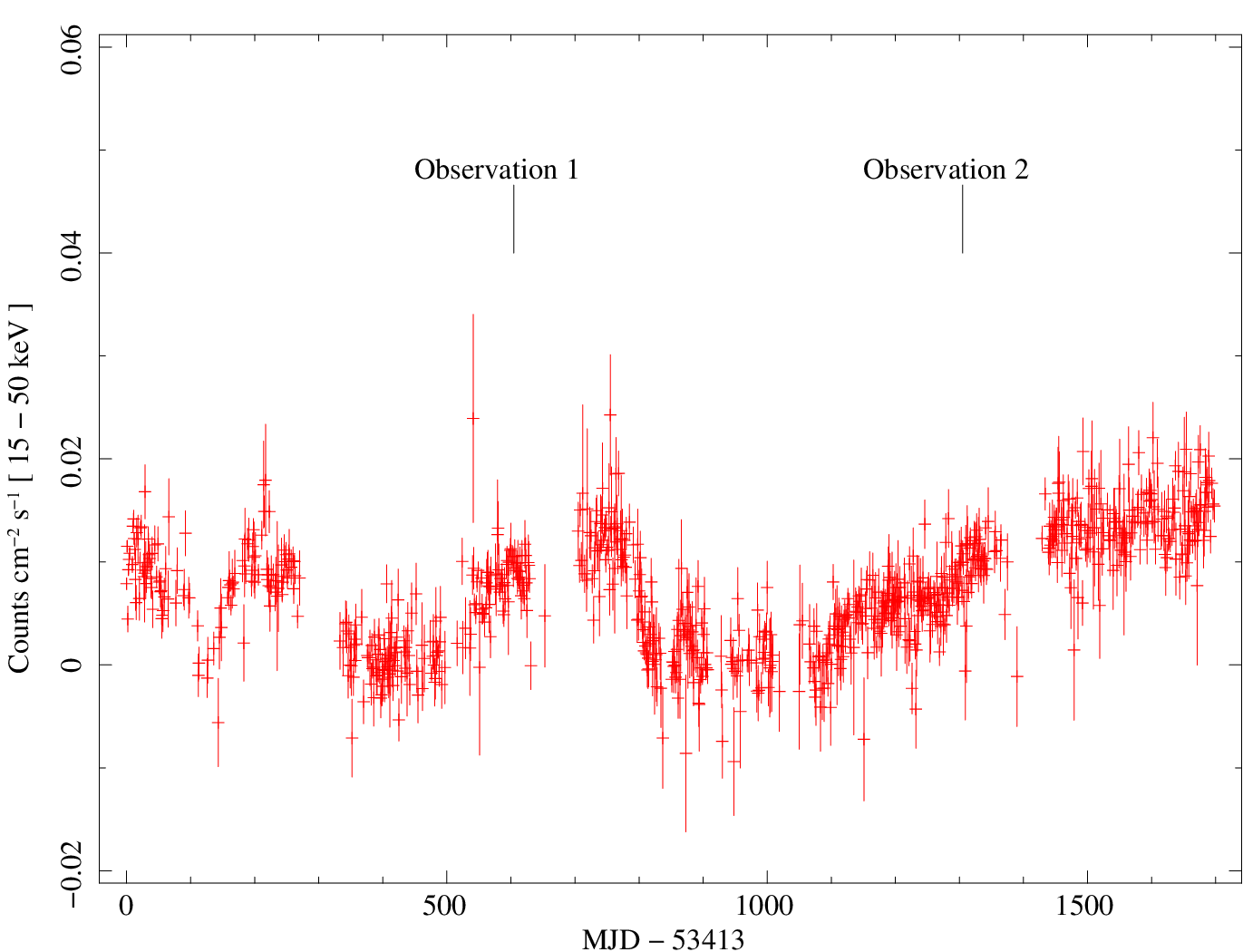}
\caption{{\it Swift} BAT hard X-ray lightcurve for \mq~from May 2005 to Jan
  2009, only exposures greater than 1ks are plotted. The times of the
  \textit{Suzaku} observations presented in this paper are indicted.}
\label{asmlc}
\end{center}
\end{figure}

\mq~was observed on two separate occasions while in the low-hard state by
\textit{Suzaku} \citep{a9} from 2006 October 09 02:20 UT until October 09
13:39 UT (obsid:501050010, PI: Koyama, epoch I) and from 2008
September 08 09:08 UT until September 09 21:33 UT (obsid:503011010, PI:
Koyama, epoch II), see Fig. \ref{asmlc}. Data were acquired over a
broad spectral range (0.2 -- 600 keV), with the X-ray imaging spectrometer
(XIS: \citealt{a10}) and the hard X-ray detector (HXD: \citealt{a11,a12}).
The source was observed at the XIS nominal position for total uncorrected
exposure times of $\sim$ 22 ks \& 18 ks (epoch I) and $\sim$ 58 ks \& 37 ks
(epoch II) for the XIS and HXD detectors respectively .

All data reduction and analysis takes place within the \textsc{heasoft
6.6.1} environment, which includes \textsc{ftools 6.6, suzaku 11} and
\textsc{xspec 12.5.0}. The latest versions of the relevant \textit{Suzaku}
\textsc{caldb} files were also used.

\subsection{X-ray Imaging Spectrometer}
The XIS has a field of view of $\sim$ 18' x 18' (1024$^2$ pixels) and was
operated in 5x5 and 3x3 readout mode. In addition the data was taken in full
window mode providing a time resolution of 8 seconds. The raw data were
processed following standard procedures, for details see \textit{Suzaku} abc
guide\footnote{
  http://heasarc.gsfc.nasa.gov/docs/suzaku/analysis/abc/}. Photon pile-up
was not an issue during these observations, so a circular extraction region
of radius 250 pixels was utilized (250 pixels is the recommended radius to
extract $\sim$ 99\% of a point source flux). In both observations \mq~is
located close to the edge of the detector, hence our extraction region is
the intersection of the detector edge and the circular aperture. The
resultant extraction region has an effective area equivalent to $\sim$ 95 \%
of the 250 pixel radius, and hence we expect to have detected a commensurate
percentage of the total source flux in the soft X-ray band ($<$ 10 keV).
Background spectra were extracted from a neighbouring region of the
detector, Response files were generated using the tasks \textsc{xisrmfgen}
and \textsc{xissimarfgen}. The background and response files were then
grouped with the science spectrum for analysis in \textsc{xspec}.

\subsection{Hard X-Ray Detector}
The HXD covers the energy range from 10 -- 600 keV, consisting of two
separate detectors, (i) PIN: Silicon PIN photodiodes covering the energy
range 10 -- 70 keV and, (ii) GSO: GSO/BGO phoswich scintillators covering
the energy range 40 - 600 keV. This instrument has a 35'$\times$35' field of
view (FoV) at energies below 100 keV, while above this the FoV is $\sim$
4.5$^{\degr}\times$4.5$^{\degr}$.  As the time of writing, the GSO data
processing routines had not been included in the official \textit{Suzaku}
data reduction pipeline, all of the hard X-ray detector data was reprocessed
following the prescription in the abc-guide$\rm ^2$, see also \citet{b16}
for a detailed description of the PIN/GSO extraction procedure.

\begin{figure*}
\begin{center}
\includegraphics[height=0.48\textheight,width=0.42\textwidth,angle=-90]{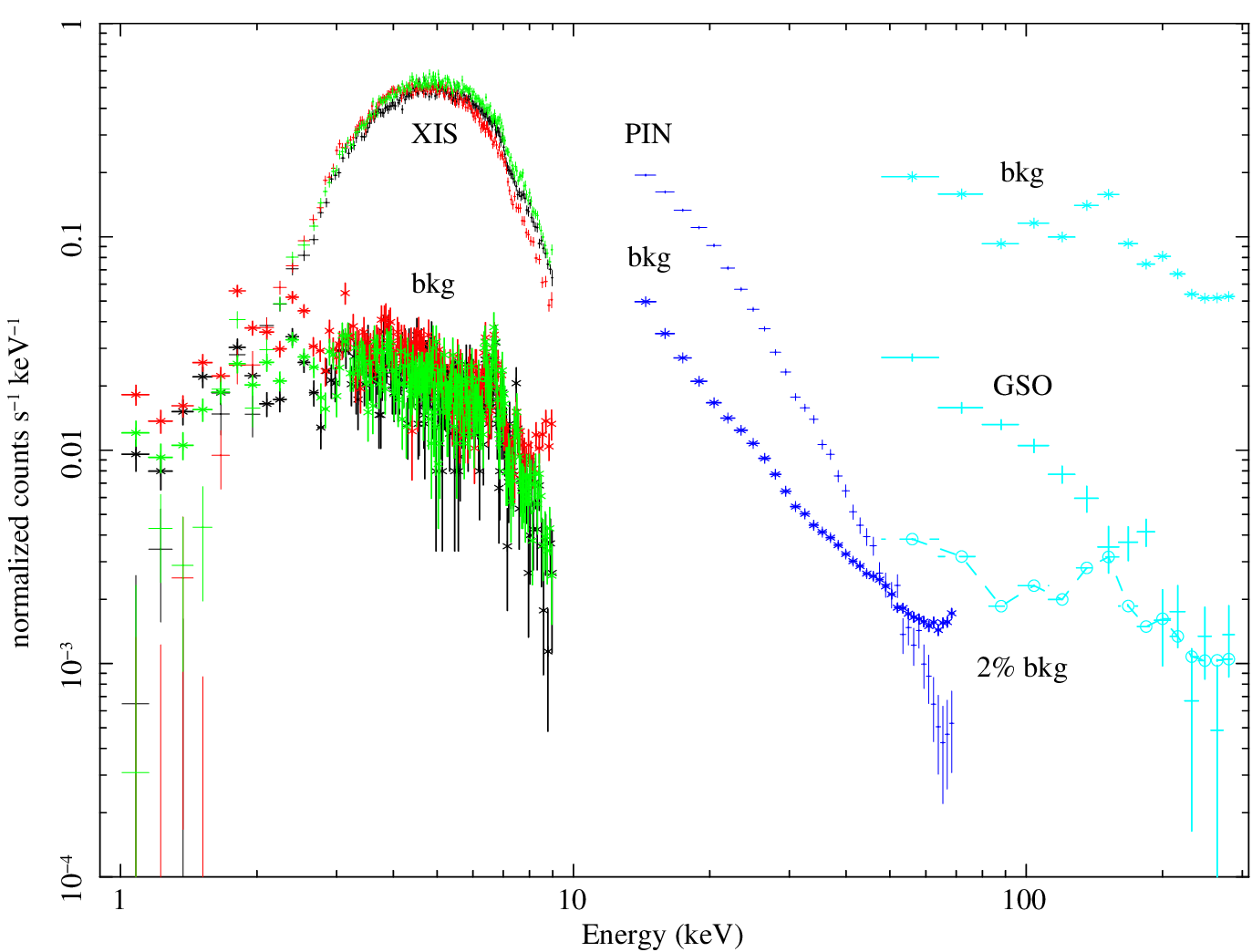}
\caption{Background subtracted XIS, PIN and GSO spectra and their
  associated background spectra. \mq~is detected between 2 -- 300 keV, with
  the low energy cutoff due to the high column density towards this
  source. The XIS detects the source between energies of 2 -- 10
  keV, the PIN detects it out to 70 keV, while the GSO detects flux to
  $\sim$ 300 keV assuming the background is reproducible to an accuracy of
  $<$2\% (see text).} 
\label{spec_back}
\end{center}
\end{figure*}

\section{Analysis \& Results}

\subsection{Soft X-ray Background}\label{backgrounds}
X-ray observations of the Galactic center have revealed a diffuse X-ray
emission component, the Galactic center diffuse X-ray emission (GCDX,
e.g. \citealt{b21, b20}). Observations with \textit{Suzaku} have revealed
this emission to be consistent with emission from a collisionally ionized
plasma with a temperature of 5 -- 7 keV.  A number of distinct atomic
emission lines have been observed from this plasma. In particular, emission
from the Iron lines Fe I K$\alpha$, Fe XXV K$\alpha$, Fe XXVI K$\alpha$ has
been resolved \citep{b20}. As such, it is important that we account for any
contamination from the GCDX in our \mq~spectrum.

\begin{figure*}
\begin{center}
\includegraphics[height=0.48\textheight,width=0.42\textwidth,angle=-90]{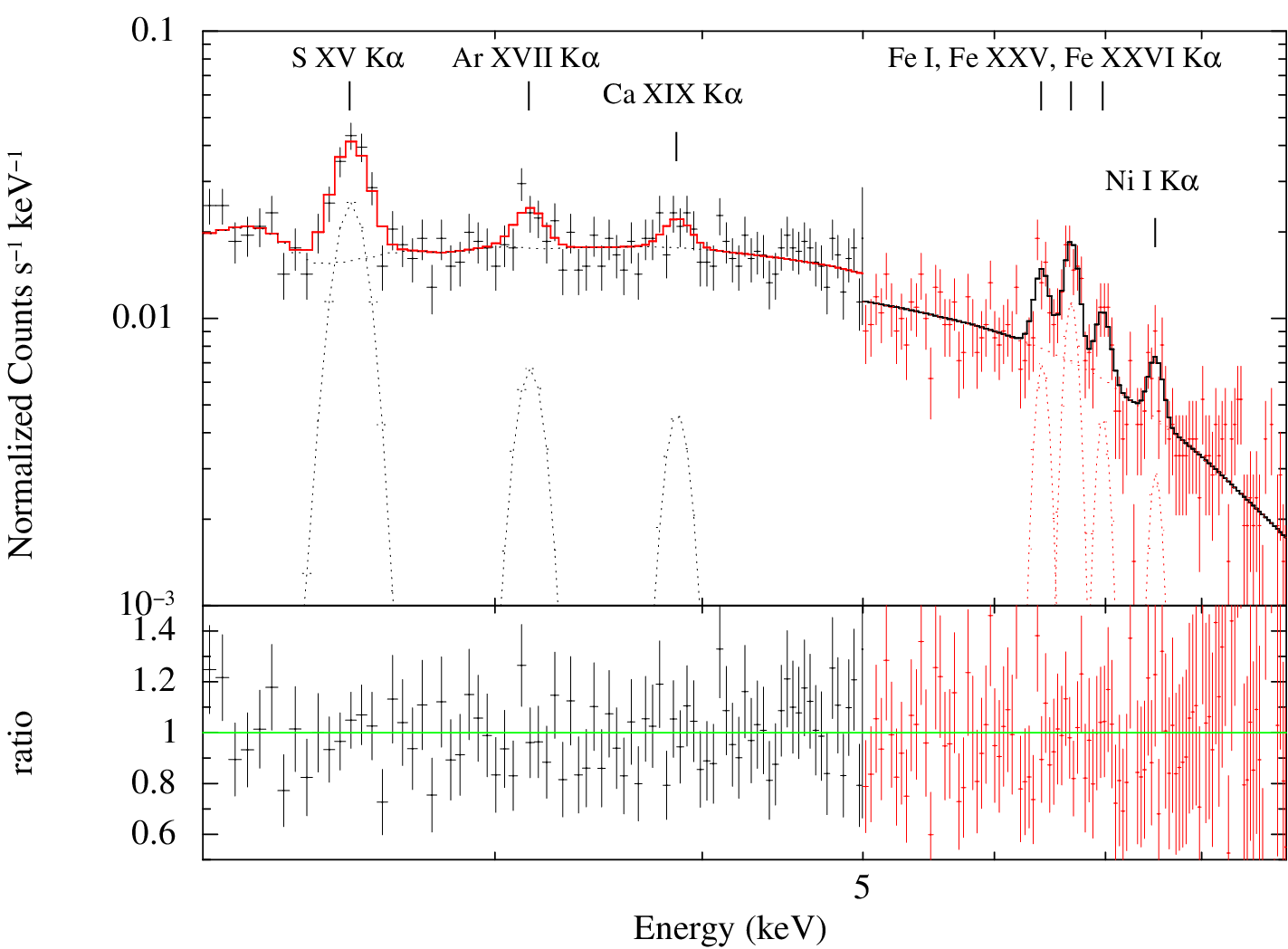}
\caption{Spectrum of the soft X-ray background towards \mq. XIS1 data is
  displayed below 5 keV (black), with data from XIS3 displayed above 5 keV
  (red). A number of prominent emission lines are clearly detected (see text
  \& Table \ref{specfit_params2}).}
\label{xis_back_spec}
\end{center}
\end{figure*}

In Fig. \ref{xis_back_spec}, we plot the background spectrum extracted from
the \mq~epoch II spectra. Here due to the different sensitivities of the front
illuminated (XIS1) and back illuminated (XIS0, XIS3) detectors in the XIS, we
use the data from XIS1 for the low energy band ($<$ 5 keV) and the data from
XIS0, XIS3 above this. A number of lines from the GCDX are clearly detected,
the line parameters are listed in Table \ref{specfit_params2}. 
The ratio of the fluxes for the Fe-lines are consistent with those measured
towards the Galactic center \citep{b22}.

\subsection{Broadband Spectra}\label{spec_i}
As the extinction is high, we do not detect any significant flux below 2 keV
in either epoch. In Fig. \ref{spec_back}, we plot the spectra extracted from
the epoch II data along with their associated background spectra. Assuming
that the systematics in the GSO background are understood at the $<$2\%
level \citep{b17}, \mq~is detected out to $\sim$ 300 keV. For reference, at
a systematic level of 3\% the flux from \mq~is equal to that of the
background at 300 keV.  

\mq~resides in a crowded region of the Galactic plane; as such, it is
important to check for contamination from nearby
sources. Fortuitously, the Galactic center was, and continues to be,
observed by \textit{INTEGRAL}, as part of the Galactic bulge monitoring
program\footnote{http://isdc.unige.ch/Science/BULGE/},
e.g. \citealt{b61}. This area was observed within a couple of days of the
\textit{Suzaku} observations presented in this paper (Revolution \#488,
\#721), and these observations provide us with flux measurements for other
sources that were bright at hard X-ray energies during our observations.

On both occasions \mq~was the dominant bright hard X-ray source in this
field. The nearest bright hard X-ray source, 1A 1742-294, was active during
the first observation only. However, this system lies 31' from \mq, which
was approximately centered in the HXD FoV, and hence lay outside
our FoV during the observations presented in this paper. As such, below 100
keV we only detect flux from \mq. We are unable to rule out the presence of
contaminating flux above this energy as there are a number of hard X-ray
sources in the large 4.5$^{\degr}\times$4.5$^{\degr}$ FoV. We note that the
consistency between the model normalizations for the PIN \& GSO spectra
would argue against the presence of a significant contaminating flux above
100 keV.

Our final good spectral range is XIS0, XIS3: 2 - 10 keV, XIS1: 2 - 10 keV,
PIN: 12 - 70 keV, and GSO: 50 - 300 keV. Additionally the data were rebinned
prior to fitting in \textsc{xspec} with the ftool \textsc{grppha} by a
factors of 4 and 8 for the XIS/PIN \& GSO spectra respectively.  As the S/N
in the second epoch is higher, we focus on these spectra. The results for
the fits to the data from epoch I, which are similar to those from epoch II,
are presented in Table \ref{specfit_params1}.

\subsubsection{Power-law models} 
The resultant spectrum (2 - 300 keV) was initially fit with a model
consisting of a power-law modified by interstellar extinction (i.e. {\tt
  pha*po}). The fit is found to be a poor one ($\chi^2_\nu \sim 1.45$) with
large residuals present at higher energies.  To account for these residuals,
we considered 2 additional power-law models; a model with a cutoff at high
energies ({\tt cutoffpl}) and a broken power-law ({\tt bknpo}). The {\tt
  cutoffpl} provides a significant improvement over the power-law model
alone $\chi^2_\nu \sim 1.16$, with the cutoff energy found to be $\sim$ 150
keV (epoch II).  The broken power-law provides the best fit to the data
$\chi^2_\nu \sim 1.04$, with the spectrum observed to break from a very hard
power-law, $\Gamma \sim$ 1.4, to a softer power-law, $\Gamma \sim$ 1.8, at
an energy of approximately 7 keV. This clearly points towards a second
component at low energies in the observed spectrum, with the most obvious
culprit being thermal emission from a geometrically thin optically thick
accretion disc \citep{b52}, see Table \ref{specfit_params1} for the detailed
fit model parameters.

Inspection of the residuals from this model reveals a number of features in
the XIS spectral range. At low energies (2 -- 3 keV), there remains a
feature consistent with the expected position of the S XV line (2.45 keV),
in addition to the presence of known systematic features in the 2.1 -- 2.4
keV region. At energies 9 - 10 keV there is an unidentified residual (in the
form of an excess possibly due to the background although its actual origin
is unclear), as such in all further modelling both of these regions are
ignored.

A large residual also exists at energies above the iron K absorption edge (E
= 7.11 keV). It is clear that we must accurately account for this spectral
feature. To do this we use the variable iron abundance absorption model
available in {\sc xspec} -- {\tt zvfeabs}. The absorption edge energy was
frozen at 7.11 keV (allowing it to vary does not significantly improve the
fit). As there exists a degeneracy in this model between the Hydrogen column
density and the metal/iron abundance, $\rm N_H$ was frozen at a value of
$\rm 1\times 10^{23}~cm^{-2}$ in agreement with the previously determined
\textit{Chandra} value \citep{b10}.
 
A blackbody accretion disc was now added to the power-law models above
(e.g. {\tt zvfeabs*(diskbb+po)}) and the fitting was repeated. The resulting
fit is significantly improved and is equal to the broken power-law model,
see Table \ref{specfit_params1}. In this case there is no difference between
the cutoff power-law and power-law models. However, the cutoff energy for
both epochs pegs at the {\sc xspec} hard limit of 500 keV, formally we find
a lower limit for the cutoff energy of $\sim$ 340 keV \& $\sim$ 390 keV at
the 90\% confidence level in epochs I \& II respectively.  It is rare that
spectral breaks are restricted to such high energies in the low-hard state
and this may be related to the fact that some models require only a small
seed photon fraction to fit the data.  The accretion disc is found to have a
temperature of $\sim$ 0.4 keV, while the disc normalization points towards a
small inner radius for the accretion disc. The metal abundance is found to
be high in agreement with previous \textit{ASCA} observations where the iron
abundance was measured to be $\sim$ 2x solar \citep{b25}.  We find an iron
abundance of $\rm N_{Fe} = 1.7\pm0.15$ and a metals abundance of $\rm N_{Z}
= 1.95\pm0.15$ relative to solar abundances.

The disc normalization may be used to estimate the inner radius of the
accretion disc when knowledge of the distance and inclination are available,
as norm $\rm \sim (r_{in}/d_{10~kpc})^2 cos\theta$. This estimate is
subject to a number of well known corrections, e.g spectral hardening
\citep{a47}, inner radius \citep{a61} and a number of others which are
difficult to quantify, e.g. zero torque inner boundary condition \citep{a46}
and radiative transfer effects \citep{a35}. Assuming a distance of 8.5 kpc,
we measure an inner radius of $\sim$ 250 (cos{\it i}$\rm ^{-1} d_{8.5kpc}$)
km and $\sim$ 120 (cos{\it i}$\rm ^{-1} d_{8.5kpc}$) km for epochs I \& II
respectively assuming a power-law continuum (see Table
\ref{specfit_params1}). We note that the ISCO for a Schwarzschild (Kerr)
black hole is 6 (1.23) $\rm R_g$, for a 10 M$_{\sun}$ black hole this is
equivalent to a radius of 90 (15) km. The inclination of \mq~ is
unconstrained; however, the presence of a bi-polar radio jet \citep{b4} would
suggest a high inclination. For instance an inclination angle of
70$^{\degr}$, will result in an increase in the inner radii above by a
factor of 3. Nonetheless the inner radii measured here are consistent with
being close to the ISCO, i.e. $\rm R_{in} \lesssim 20~R_g$.

We test for the presence of a broad iron line by adding a {\tt laor}
component to the model, whose energy is restricted to the 6.4 -- 6.97 keV
range (q = 3, $\rm R_{out}$ = 400, i = 70$^{\degr}$). The presence of a
broad line ($\rm E \sim 6.7 keV, R_{in} \sim 3~R_g$) is only marginally
required (99\% confidence level as measured with an F-test). As such we
freeze the inner radius at 3 $\rm R_g$ and step the line energy through the
energy region outlined above. For a relativistic line centered at 6.4 keV,
we can place an upper limit in the EW of this line of $\sim$ 60 eV, for
higher energies, e.g. 6.7, 6.9 keV the upper limit increases to $\sim$ 150,
200 eV respectively.

Likewise in order to investigate the possible contribution due to disc
reflection at higher energies (20 -- 40 keV), the best fit power-law from
above was convolved with the reflection model of \citet{a43} {\tt
  reflect*po}. This model was then relativistically blurred ({\tt kdblur})
to account for the proximity of the accretion disc to the ISCO as indicated
above.  The inclination was held fixed (cos{\it i} = 0.45), while the
abundances of metals were tied to those of the absorption component. Any
reflection fraction is negligible (f $\sim$ 1\%). Formally, we find an upper
limit for disc reflection of 10\% at the 90\% confidence level.

\citet{b13} have detected numerous absorption lines, in particular Fe XXV,
XXVI , attributed to a disc wind in soft state observations of the
microquasar GRS 1915+105. Earlier hard state observations by \citet{b14} had
also detected some of these lines; however, they were significantly weaker.
Here we test for the presence of narrow Gaussian absorption lines in the Fe
region of the spectrum. No significant absorption lines are detected, at the
90\% confidence level we place an upper limit on the equivalent width of any
absorption line of $<$ 3 eV.

We note that due to the higher S/N in the second epochs data, this fit
provides tighter spectral constraints, which nonetheless are consistent with
those from epoch I within the errors.

\subsubsection{Comptonization models}
To investigate the Comptonizing medium/region, we use the bulk motion
Comptonization model {-- \tt bmc} \citep{b19}. This is a generic
Comptonization model which includes the flux from the accretion disc and the
fraction of this seed flux Comptonized in the corona. We use this model as
it considers the general case where the Comptonization process may be either
thermal or dynamic in nature\footnote{In the thermal Comptonization case,
  for energies less than $\rm \sim 2kT_e$, the {\tt comptt} \& {\tt bmc}
  models are similar.}. In the {\tt bmc} model the soft photons are those
that undergo few scatterings, while those in the hard component have
undergone multiple scatterings.

\begin{figure*}
\begin{center}
\includegraphics[height=0.48\textheight,width=0.42\textwidth,angle=-90]{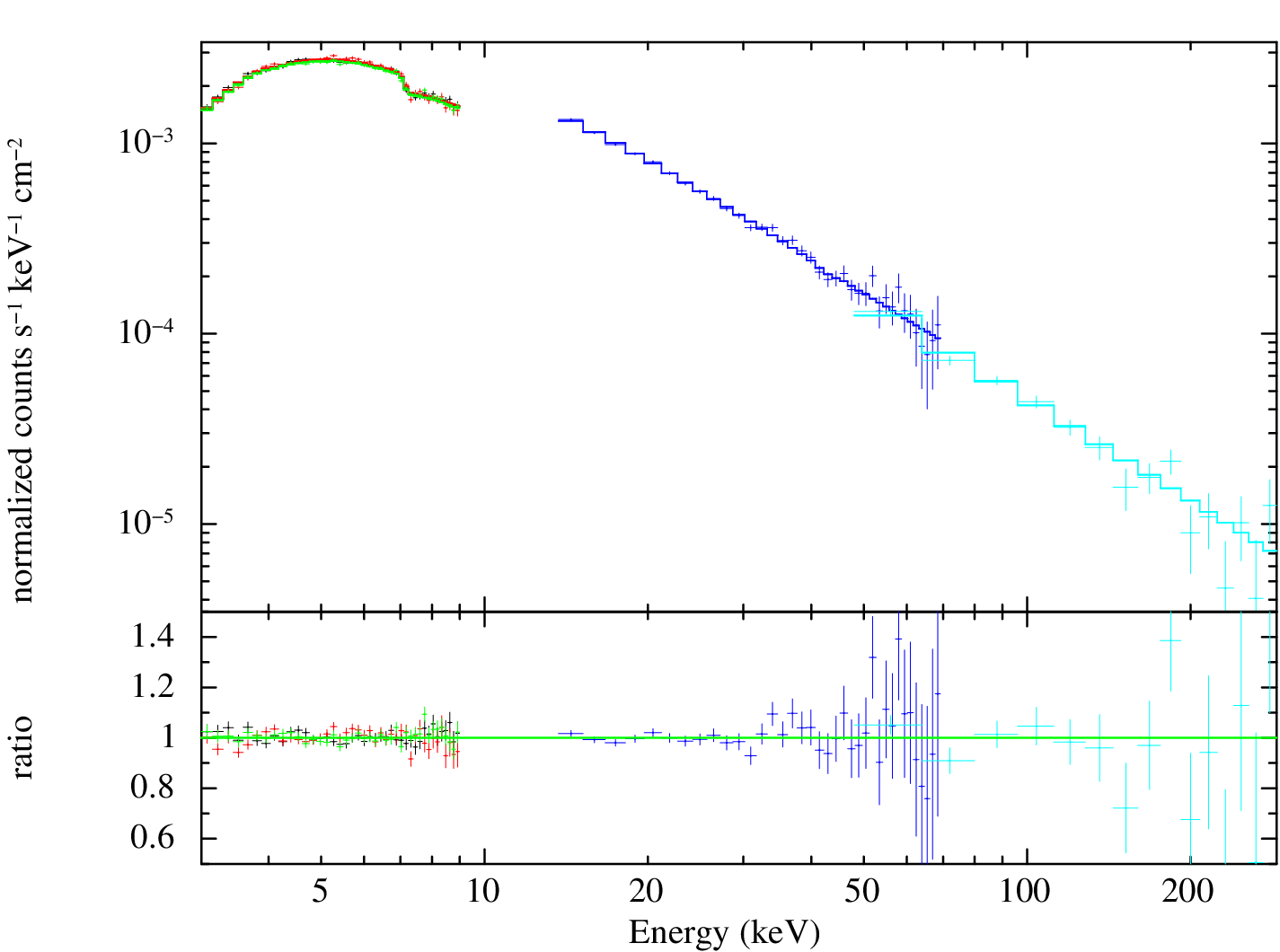}
\caption{Best fit to the epoch II spectrum, consisting of Comptonization
  modified by absorption with a variable iron abundance, {\tt zvfeabs*bmc}:
  $\rm \chi^2_{\nu} \sim 1.04$. The thermal component temperature is $\rm
  \sim 0.36~keV$, and the photon index is $\sim 1.8$, see Table 2. The XIS0,
  XIS1, XIS2 spectra are indicated by the black, red and green points
  respectively, while the PIN (blue) and GSO (light blue) are also
  plotted.}
\label{spec_bmc}
\end{center}
\end{figure*}

Our fit results in a reduced chi-squared of $\sim$ 1.04 (1327/1272 -- epoch
II, 1297/1272 -- epoch I, see Fig. \ref{spec_bmc}), with the best fit
parameters consistent with those found in the previous section. The
temperature of the soft disc component for epoch II (I) is 0.36$\pm$0.02 keV
(0.28$\pm$0.06 keV). The spectral index $\alpha$ agrees with our
expectations from the {\tt diskbb+po} fit, i.e. $\rm \Gamma = 1.79\pm0.03$
($\rm 1.82\pm0.04$) where $\rm \alpha = \Gamma - 1$, while $\rm A =
-0.62\pm0.09$ ($-0.82^{+0.46}_{-0.17}$).  We find both iron and metals to be
over abundant relative to solar ($\sim$ 2x, 1.7x respectively).

\begin{table}
\begin{center}
\caption{2 - 10 keV Background Features \label{specfit_params2}}
\begin{tabular}{lcc}
\tableline\\ [-2.0ex]
 K$\alpha$ Line & Energy & EW \\
 & [ keV ] & [ eV ]\\ [0.5ex]
\tableline\tableline\\ [-2.0ex] 
S XV  &  2.45$\pm$0.01 & 240$^{+400}_{-100}$\\ [0.5ex]   
Ar XVII  &  3.15$\pm$0.03 & 65$^{+250}$\\ [0.5ex]
Ca XIX &  3.86$\pm$0.06 & 48$^{+70}$\\  [0.5ex] 
Fe I &  6.41$\pm$0.02 & 100 $^{+200}_{-70}$\\   [0.5ex] 
Fe XXV &  6.68$\pm$0.02 & 200 $^{+200}_{-130}$\\  [0.5ex] 
Fe XXVI &  6.98$\pm$0.3 & 90 $^{+200}$\\ [0.5ex]
Ni I &  7.51$\pm$0.05 & 120$^{+200}$ \\  [0.5ex]  
\tableline\\ [-2.0ex]
\end{tabular}
\tablecomments{Atomic lines detected in the \mq~background spectrum. All
  errors are quoted at the 90\% confidence level. Note: where lower limits
  are not given, the lower limit is consistent with zero.}
\end{center}
\end{table}

Testing for the presence of a relativistic iron line reveals results
consistent with those in the previous section.  We also attempted to
constrain the possible presence of non-thermal electrons at higher energies
using the hybrid Comptonization model {\tt eqpair} \citep{b51}; however, we
are unable to constrain the fraction of non-thermal/thermal electrons. As
discussed in \citet{b51}, this is unsurprising as in order to constrain the
non-thermal fraction, one requires a source detection out to energies
approaching 500 keV.

\begin{table*}
\begin{center}
\caption{Spectral Fit Parameters \label{specfit_params1}}
\begin{tabular}{lcccccccccc}
\tableline\\ [-2.0ex]
Model & Epoch & $\rm N_{Z}$ & $\rm N_{Fe}$ & $\rm \Gamma_1$ & $\Gamma_2$ &
$\rm E_{cut}$ & $\rm E_{break}$ & & $\chi^2/\nu$ \\
 & & [ $\rm N_{z_{\sun}}$ ] & [ $\rm N_{Fe_{\sun}}$ ] & & & [ $\rm keV$ ] &
[ $\rm keV$ ] & & \\ [0.5ex] 
\tableline\\ [-2.0ex] 
{\tt po} & I & 1.36$\pm$0.06 & 1.55$\pm$0.20 & 1.74$\pm$0.04 & -- & -- & -- & -- & 1335.91/1274 \\ [0.5ex]
{\tt po} & II & 1.44$\pm$0.04 & 1.29$\pm$0.13 & 1.67$\pm$0.02 & -- & -- & --
& -- & 1538.04/1274 \\ \\[0.5ex]
{\tt cutoffpl} & I &  1.31$^{+0.07}_{-0.08}$ & 1.63$\pm$0.22 &
1.68$^{+0.04}_{-0.07}$ & -- & -- & 335$^{> 500}_{-160}$ & -- & 1332.85/1273 \\ [0.5ex]
{\tt cutoffpl} & II &  1.30$\pm$0.05 & 1.51$\pm$0.14 & 1.52$\pm$0.04 & -- &
-- & 143$^{+44}_{-28}$ & -- & 1473.81/1269 \\ \\[0.5ex]
{\tt bknpo} & I & 1.06$\pm$0.10 & 1.34$\pm$0.22 & 1.27$^{+0.13}_{-0.08}$ &
1.79$\pm$0.04 & -- & 6.0$\pm$0.3 & -- & 1293.43/1272 \\[0.5ex]
{\tt bknpo} & II & 0.99$\pm$0.06 & 1.99$\pm$0.15 & 1.29$\pm$0.05 &
1.79$\pm$0.03 & -- & 7.7$\pm$0.2        & -- & 1327.99/1272 \\ \\ [0.5ex]
\tableline\\ [-2.0ex]
Model & Epoch & $\rm N_Z$ & $\rm N_Z$ & $\rm T_{bb}$ & $\rm XIS1_{norm}$ &
$\rm R_{in}$ & $\rm \Gamma$ & $\rm E_{cut}$ & $\chi^2/\nu$ \\
 & &  [ $\rm N_{Z_{\sun}}$ ] & $\rm N_{Fe_{\sun}}$ & [ $\rm keV$ ] & & [
  $\rm km$ ] & [ $\rm keV$ ] & [ $\rm keV$ ] & \\ [0.5ex]
\tableline\\ [-2.0ex]
{\tt diskbb+po} & I & 1.58$\pm$0.11 & 1.67$\pm$0.22 & 0.32$\pm$0.08 &
21230$^{+7.1e5}_{-18219}$& 248$^{+1205}_{-154}$ & 1.82$\pm$0.04 & -- & 1292/1270 \\ [0.5ex]
{\tt diskbb+po} & II & 1.98$\pm$0.10 & 1.70$\pm$0.15 & 0.41$\pm$0.02 &
4696$^{+2206}_{-1442}$& 117$^{+24}_{-20}$ & 1.77$\pm$0.03 & -- & 1317/1270 \\  [0.5ex]
{\tt diskbb+cutoffpl} & I & 1.51$^{+0.12}_{0.10}$ & 1.70$\pm$0.22 &
0.30$\pm$0.09 & 26707$^{+1.9e6}_{-24111}$ & 278$^{+2082}_{-190}$ & 1.77$\pm$0.04 & $>$ 343 & 1296/1269 \\  [0.5ex]
{\tt diskbb+cutoffpl} & II & 1.89$\pm$0.10 & 1.72$\pm$0.15 & 0.41$\pm$0.02 &
3537$^{+1937}_{-1165}$ & 101$^{+25}_{-18}$ & 1.74$\pm$0.03 & $>$ 389 & 1320/1269 \\ [0.5ex]
\tableline\\ [-2.0ex]
Model & Epoch & $\rm N_{Z}$ & $\rm N_{Fe}$ & kT & $\rm \alpha$ & log(A) & 
& &  $\chi^2/\nu$ \\
 & & [ $\rm N_{z_{\sun}}$ ] & [ $\rm N_{Fe_{\sun}}$ ] & [$\rm keV$ ] & & & & \\ [0.5ex]
\tableline\\ [-2.0ex] 
{\tt bmc} & I & 1.58$^{+0.12}_{-0.10}$ & 1.67$\pm$0.21 &
0.28$^{+0.06}_{-0.04}$ & 0.82$\pm$0.04 & -0.82$^{+0.46}_{-0.99}$  & -- & -- &  1297/1272 \\ [0.5ex] 
{\tt bmc} & II & 1.95$\pm$0.1 & 1.69$\pm$0.15 & 0.36$\pm$0.02 &
0.79$\pm$0.03 & -0.62$\pm$0.09 & -- & -- & 1327/1272\\ [0.5ex]  
\tableline\\ [-2.0ex]
\end{tabular}
\tablecomments{The Hydrogen column density was frozen at a value of $\rm 1
  \times 10^{23}~cm^{-2}$ in agreement with previous estimates
  \citep{b10}. The iron absorption edge energy was also fixed at 7.11
  keV. Errors quoted are 90\% confidence intervals determined via the {\tt
    error} command in {\sc xspec}. The accretion disc inner radius is
  calculated assuming a distance of 8.5 kpc, additionally the inclination
  dependence of (cos{\it i}$^{-1}$) has not been included.} 
\end{center}
\end{table*}

\section{Discussion}
We present \textit{Suzaku} broadband spectra of the Galactic microquasar
\mq~while in the low-hard state. The source was observed on two separate
occasions after a transition from the high-soft state to the low-hard state
at a luminosity of $\sim 1\%$ Eddington, i.e. a 2 - 300 keV unabsorbed flux
of $\rm 2.2 \times 10^{-9}~erg~s^{-1}~cm^{-2}$ ($\rm L_x/L_{Edd} = 0.014
\times (d/8.5~kpc)^2(M_x/10~M_{\sun})$).

Given the large background at energies above 200 keV, it is worth
investigating the effect of ignoring data beyond this, and how this might
affect the presence of a spectral cut-off in our data. Ignoring all data in
excess of 200 keV and refitting the higher quality epoch II data does not
significantly change our results, with a lower limit to the spectral cutoff
of $\sim$ 375 keV (90\% confidence) in this case. We note, that the best fit
cut-off power-law model, where the spectral cut-off is fixed at an energy of
200 keV, is excluded at the $\rm > 3\sigma$ \& $\rm > 4\sigma$ level, as
measured via a F-test, for epochs I \& II respectively, see Table
\ref{specfit_params1}

The observed low-hard state spectrum of \mq~as presented herein, and via the
\textit{INTEGRAL} observations presented by \citet{b8}, is clearly
consistent with a power-law.  Here \mq~is detected up to an energy of 300
keV.  A prominent iron K absorption edge is visible in the XIS soft X-ray
spectra. We account for this with the {\tt zvfeabs} model in {\sc xspec} and
find the metal \& iron abundances to be super-solar for all our models in
agreement with previous \textit{ASCA} observations \citep{b25}. Modeling the
spectrum as the sum of a blackbody accretion disc and an unbroken power-law
at higher energies provides a good fit, although we can not rule out the
presence of a spectral cut off at energies $\gtrsim$ 380 keV.

The accretion disc temperature is determined to be low ($\sim$ 0.4 keV),
while we find the disc inner radius to be small and consistent with a disc
that is continuous down to small radii approaching the ISCO (i.e. $\rm
R_{in} \lesssim 20~R_g$; see Table \ref{specfit_params1}). A broad iron line
is also required but at a low significance (99\% as measured by an F-test),
supporting a small inner radius for the accretion disc.  The detection of a
cool disk in \mq~is consistent with a growing body of evidence that the
accretion disk is not necessarily truncated at large radii in the low-hard
state.

The spectrum is characterized in detail using the {\tt bmc} Comptonization
model \citep{b19}. A cool source of thermal photons is revealed in this
model consistent with the presence of a cool accretion disc, as suggested by
the phenomenological modelling above. The parameter log(A) in the {\tt bmc}
model is related to the fraction of Comptonized soft seed photons $\rm f =
A/1+A$, in these observations the fraction is low.  In epoch II the
Comptonized fraction is $\sim$ 20\% while in epoch I it is lower still at
15\%. Unfortunately the lower S/N in epoch I prevents us from making any
definitive statements regarding the evolution of the coronal geometry
between observations.

\subsection{Previous Observations}
\citet{b8} have observed this system with \textit{INTEGRAL} in the hard
state, and detect it out to an energy of 600 keV. The spectrum is found to
be consistent with thermal Comptonization up to 200 keV with an additional
component required at energies above this. They fit the spectrum with a two
component thermal Comptonization model ($\rm kT_{e1} \sim 30~keV,~kT_{e2}
\sim 100~keV$) although they are unable to rule out non-thermal processes.
Indeed fitting the spectrum above 200 keV with a power-law produces a high
quality fit ($\chi^2_\nu \sim 1$). Such a power-law could be produced by
non-thermal processes such as those expected from a jet, e.g. \citet{b54}.
The measurement of a power-law spectrum for \mq~out to 600 keV by
\textit{INTEGRAL} is consistent with the \textit{Suzaku} observations
presented herein.

\citet{b9} carried out a series of simultaneous \textit{RXTE} \&
\textit{INTEGRAL} observations primarily while \mq~was in the low-hard
state. The spectrum was found to be consistent with a power-law ($\rm \Gamma
\sim$ 1.3 -- 1.6) including a high energy cut-off ($\sim$ 100 -- 120 keV);
however, we note that the S/N of the high energy ($>$ 50 keV) portion of the
spectrum is very low, with only a single data bin at $>$ 100 keV. A
significant amount of reflection from the accretion disc was also required,
with typical values for the reflection fraction of 0.3 -- 0.9. Fitting with
the {\tt compps} model reveals a seed photon temperature of 0.4 -- 0.7 keV,
consistent with the accretion disc temperature we measure here.

We do not observe any high energy cut-off features, as observed by
\citet{b9}, in our data. Although, we note the power-law emission is clearly
softer in our case ($\rm \Gamma \sim 1.8$).  We do not observe any evidence
for significant disc reflection (either an iron line or Compton hump) in the
spectra from either epoch, with a detection of a weak line at the 99\%
confidence level (EW $\sim$ 60 eV) in addition to a weak disk reflection
fraction (f $\lesssim$ 0.1). Given the high level of Comptonization clearly
present, it is likely that any fluorescent iron line emission produced in
the inner disc would be smeared by the corona \citep{b55}. The lack of these
features is consistent with the expectation of a radially recessed/truncated
disc in the hard state \citep{b43} and/or the presence of an outflow
\citep{a38,a39,a25}.

Observations of \mq~with \textit{Chandra} also measure a hard power-law spectrum
consistent with the expectations for the hard state $\rm \Gamma \sim 1.4$
\citep{b10}. We note that a power-law fit to our \textit{Suzaku} spectrum in
the 2 -- 10 keV range also returns a photon index of $\sim$ 1.4 suggesting
that the \textit{Chandra} spectrum contained a significant accretion disc
component, which is only revealed through the broad energy coverage of
\textit{Suzaku}.

The observations listed above are in agreement with those presented in this
paper, and point towards the X-ray emission from \mq~being dominated by the
accretion disc \& corona while the source is in the LHS. This interpretation
is supported by the work of \citep{b26} who modelled the broadband emission
(radio - GeV) from \mq. They find the observed hard X-ray flux to be
inconsistent with the measured radio flux in the case where the spectral
energy distribution is jet dominated and instead favour the corona as the
origin of the observed high energy emission.

\subsection{Implications for the low-hard state} 
Observations of the candidate black hole binary SWIFT J1753.5-0127 in the
low-hard state with \textit{Suzaku} reveal a similar picture \citep{b16},
i.e. a low temperature accretion disc component plus an unbroken power-law
extending to high energies at a luminosity of $\rm \sim 2\%~L_{Edd}$. In
this case the disc temperature is found to be $\sim$ 0.2 keV while the inner
radius is consistent with that measured for \mq. A low reflection fraction
of $\sim$ 0.2 is measured along with an unbroken power-law and a weak iron
line, all consistent with an inner radius $\lesssim$ 20 R$\rm _g$. A similar
picture is revealed through modelling the the soft disc component and the
relativistic iron line detected by \textit{XMM} \citep{a18,a66}.  The low
values measured for the reflection fraction in both \mq~and SWIFT
J1753.5-0127 are consistent with models that do not require a recessed disc
in the low-hard state, e.g. \citet{a38,a39,a25}.

Combined with the accretion disc component and relativistically broadened
iron lines observed in numerous systems, these results provide significant
evidence opposing the view that the accretion disc is truncated at large
radii ($\rm > 100~R_g$) in the low-hard state, e.g., see also
\citet{a20,a32,a74,a75,b80}.

\subsection{Comparison to other Microquasars}
GRS 1758-258 is the system most closely resembling \mq~among the Galactic
black hole candidates. However, we first consider a series of broadband
observations of the black hole binaries XTE J1650-500 and GRO J1655-40, which
displayed low-hard state spectra similar to that presented in this paper
for \mq.

XTE J1650-500 is a Galactic black hole binary that was discovered in 2001 by
\textit{RXTE} \citep{b62}. The orbital period has been measured to be $\sim$
7.6 hrs, and the mass of the black hole is $\rm M_{BH} = 4.0 -
7.3~M_{\sun}$ \citep{b63}. Observations during the initial outburst with
\textit{XMM-Newton} detected a broad skewed Fe-K$\alpha$, which modelling
revealed to be consistent with a near maximally spinning black hole ($\rm
a^* \sim 1$; \citealt{b64}). Observations at radio wavelengths revealed
significant emission consistent with a origin in a compact jet \citep{b65}.
\citet{b33} analyze \textit{BeppoSax} observations
of XTE J1650-500 with the {\tt bmc} model\footnote{\citet{b33} also include
  an extra blackbody component to account for that portion of the disc that
  is not covered by the corona, as we do not detect any photons below 2 keV
  this component is not required.} in the the low-hard and high-soft
states. The photon index in the low-hard state is similar to our
observations $\rm \Gamma \sim 1.7 - 1.8$. The soft seed component is found
to have a temperature of $\rm \sim 0.4~keV$ again similar to our
observations of \mq. In contrast to XTE J1650-500, we find a much lower
Comptonized fraction in the case of \mq~$\sim$ 20\% (versus $\sim$ 60\%),
which implies a smaller corona. In the case of XTE J1650-500, \citet{b33}
find evidence for a contraction in the size of the Corona as the microquasar
transitions from the LH to the HS state, as such one might naturally expect
the corona to be smaller when the transition progresses in the opposite
sense. 

GRO J1655-40 was discovered by the \textit{Compton Gamma Ray Observatory} in
1994 \citep{b71} and found to lie at a distance of $\sim$ 3.2 kpc
\citep{b67}. \citet{b66} measured an orbital period of $\sim$2.6 days and
the black hole mass to be $\rm M_x = 7.02\pm0.22~M_{\sun}$. Highly
relativistic radio jets were also detected, which appeared to be misaligned
with the binary orbit \citep{b73}. X-ray observations revealed a pair of
high frequency QPOs consistent with Keplerian rotation at the ISCO of a
spinning black hole \citep{b68}. Subsequent observations detected a broadly
skewed iron K$\alpha$ line, which supports a high spin for the black hole
($a \geq 0.9$; \citealt{b69,b72}). \citet{b70} presented \textit{RXTE} and
\textit{OSSE} observations in the high-soft state which revealed unbroken
power-law emission out to an energy of $\sim$ 800 keV. \citet{a67} presented
broadband (3 -- 500 keV) \textit{INTEGRAL} observations of GRO J1655-40
during the 2005 outburst. Here an observation in the low-hard state ($\rm
L_x \sim 0.015~L_{Edd}$) clearly detected unbroken power-law emission
($\Gamma \sim 1.7$) extending out to an energy of $\sim$ 500 keV, again
consistent with the observations of \mq~presented herein. This was
interpreted as evidence for a significant contribution form non-thermal
electrons (although, see also \citealt{b74}). Unfortunately, the
\textit{INTEGRAL} low energy cut-off of 3 keV does not allow any constraints
to be placed on emission from the accretion disc.

GRS 1758-258 is the second persistent microquasar to lie close to the
Galactic center. It is similar to \mq, i.e. large scale radio outflows
\citep{b79}, X-ray bright with a high extinction \citep{b78}, no identified
optical/NIR counterpart \citep{b7} and a long orbital period ($\sim$
18.5 days; \citealt{b11}).  Detailed studies of the LHS state properties of
this system have been carried out by \citet{b28} and \citet{b29} who
observed with \textit{INTEGRAL/RXTE} \& \textit{BeppoSax}
respectively. \citet{b28} detected the source in the energy range 3 -- 200
keV and found the low-hard state spectrum to be consistent with a cut-off
power-law, where $\rm \Gamma \sim 1.5 - 1.7~and~E_{cut} \sim 140 -
250~keV$. Due to contamination from GX 5-1, \citet{b29} only detected GRS
1758-258 in the energy ranges 0.1 -- 10 keV \& 40 -- 200 keV and found the
broadband spectrum to be consistent with a cut-off power-law where $\rm
\Gamma \sim 1.65, E_{cut} \sim 70~keV~and~E_{fold} \sim 180~keV$. In both
sets of observations a weak emission line consistent with Fe--K$\alpha$ was
marginally detected with an equivalent width $\rm \sim 50 - 70~eV$. New
broadband \textit{Suzaku} observations of GRS 1758-258 will be presented in
an upcoming paper (Reynolds et al., 2010 in prep.).

\section{Conclusions}
We present \textit{Suzaku} observations of the Galactic center microquasar
\mq~ in two separate epochs taken after the system had transitioned into the
low-hard state. The system is observed to be in the low-hard state at the time
of our observations with an X-ray luminosity of $\sim$ 1\% Eddington.  The
spectra in each epoch are similar, being described by a model consisting of
a soft thermal accretion disc component (T $\sim$ 0.4 keV) and the broadband
emission ($\rm > 10 keV$) is found to be characterized by an unbroken
power-law to at least 300 keV.

Consistent with growing evidence from observations of numerous systems in
the low-hard state (e.g. GX 339-4, SWIFT J1753.5-0127, XTE J1817-330, XTE
J1118+480), we also find evidence that the accretion disc in \mq~is not
truncated at large radii in the low-hard state and instead resides close to
the ISCO ($\rm R_{in} \sim 20~R_g$).\\

\acknowledgements
We thank the anonymous referee for his/her careful review and report.  This
research has made use of data obtained from the \textit{Suzaku} satellite, a
collaborative mission between the space agencies of Japan (JAXA) and the USA
(NASA). This research made extensive use of the \textit{SIMBAD} database,
operated at CDS, Strasbourg, France and NASA's Astrophysics Data System.


\vspace{1cm}
\footnotesize{This paper was typeset using a \LaTeX\ file prepared by the 
author}


\end{document}